\documentclass[aps,prl,twocolumn,superscriptaddress,amsmath,amssymb,showpacs]{revtex4-1}

\usepackage{graphicx}	
\usepackage{dcolumn}	
\usepackage{bm}		
\usepackage{verbatim} 	
\usepackage{braket}
\usepackage{xr}
\usepackage{hyperref}

\begin{document}

\title{High Dimensional Unitary Transformations and Boson Sampling on Temporal Modes using Dispersive Optics}

\author{Mihir Pant}
\email{mpant@mit.edu}
\affiliation{Dept. of Electrical Engineering and Computer Science, MIT, Cambridge, MA 02139, USA.}

\author{Dirk Englund}
\affiliation{Dept. of Electrical Engineering and Computer Science, MIT, Cambridge, MA 02139, USA.}

\begin{abstract}

We present methods which allow orders of magnitude increase in the number of modes in linear optics experiments by moving from spatial encoding to temporal encoding and using dispersion. This enables significant practical advantages for linear quantum optics and Boson Sampling experiments. Passing consecutively heralded photons through time-independent dispersion and measuring the output time of the photons is equivalent to a Boson Sampling experiment for which no efficient classical algorithm is reported, to our knowledge. With time-dependent dispersion, it is possible to implement arbitrary single-particle unitaries. Given the relatively simple requirements of these schemes, they provide a path to realizing much larger linear quantum optics experiments including post-classical Boson Sampling machines.
\end{abstract}

\maketitle

Unitary transformations on optical modes have been used to implement single particle quantum gates \cite{2007.RMP.Kok-Milburn.LOQCRev, 2009.NPhoton.Brien-Vuckovic.PhotonQuantTech}, quantum simulations \cite{2010.Science.Peruzzo-OBrien.Qwalk}, and Boson Sampling \cite{2013.TheoryofComputing.Aaronson-Arkhipov.linear_optics_complexity, 2013.Science.Spring-Walmsley.BS_photonic_chip, 2013.Science.Broome-White.BS_photonic_tunable, 2013.NPhoton.Tillmann-Walther.BS_silica_laser, 2013.NPhoton.Crespi-Sciarrino.BS_mmInterferometer, 2014.NPhoton.Carolan-Laing.BS_5photon, 2014.ArXix.Tillmann-Walther.BS_controllable_distinguishability, 2014.NPhoton.Spagnolo-Sciarrino.BS_Exp_val}. Traditionally, these transformations are implemented on spatial modes using a system of beamsplitters. However, building a large interferometer implementing such a unitary transformation is experimentally challenging and the largest number of modes so far has been 21 \cite{2014.NPhoton.Carolan-Laing.BS_5photon}. 

A particularly interesting application which requires a large number of spatial modes is Boson Sampling. Boson Sampling is the process of estimating the output photon distribution after passing multiple identical photons through a passive linear interferometer. Aaronson and Arkhipov have proposed that Boson Sampling is computationally hard for classical computers because it requires the estimation of the permanents of independent and identically distributed (iid) Gaussian matrices, a problem that is believed to reside in the $\#P$-complete complexity class \cite{2013.TheoryofComputing.Aaronson-Arkhipov.linear_optics_complexity}. The only `effective interaction' between photons in such a system is due to the bosonic statistics of these identical photons at the detectors. The need for only linear optics and photodetection could make the Boson Sampling problem easier than general quantum computing approaches with photons, which require nonlinear materials \cite{1995.PRL.Turchette-Kimble.PhaseShiftQlogic} or feed-forward schemes \cite{2001.Nature.Knill-Milburn.KLM}. Furthermore, unlike other quantum computing schemes that require on-demand sources, Boson Sampling with probabilistic but heralded input photons has been proposed to be computationally hard for a classical computer \cite{2014.PRL.Lund-Ralph.bs_gaussian}. 

\begin{figure}[h]
    \includegraphics[width=0.5\textwidth, page=1]{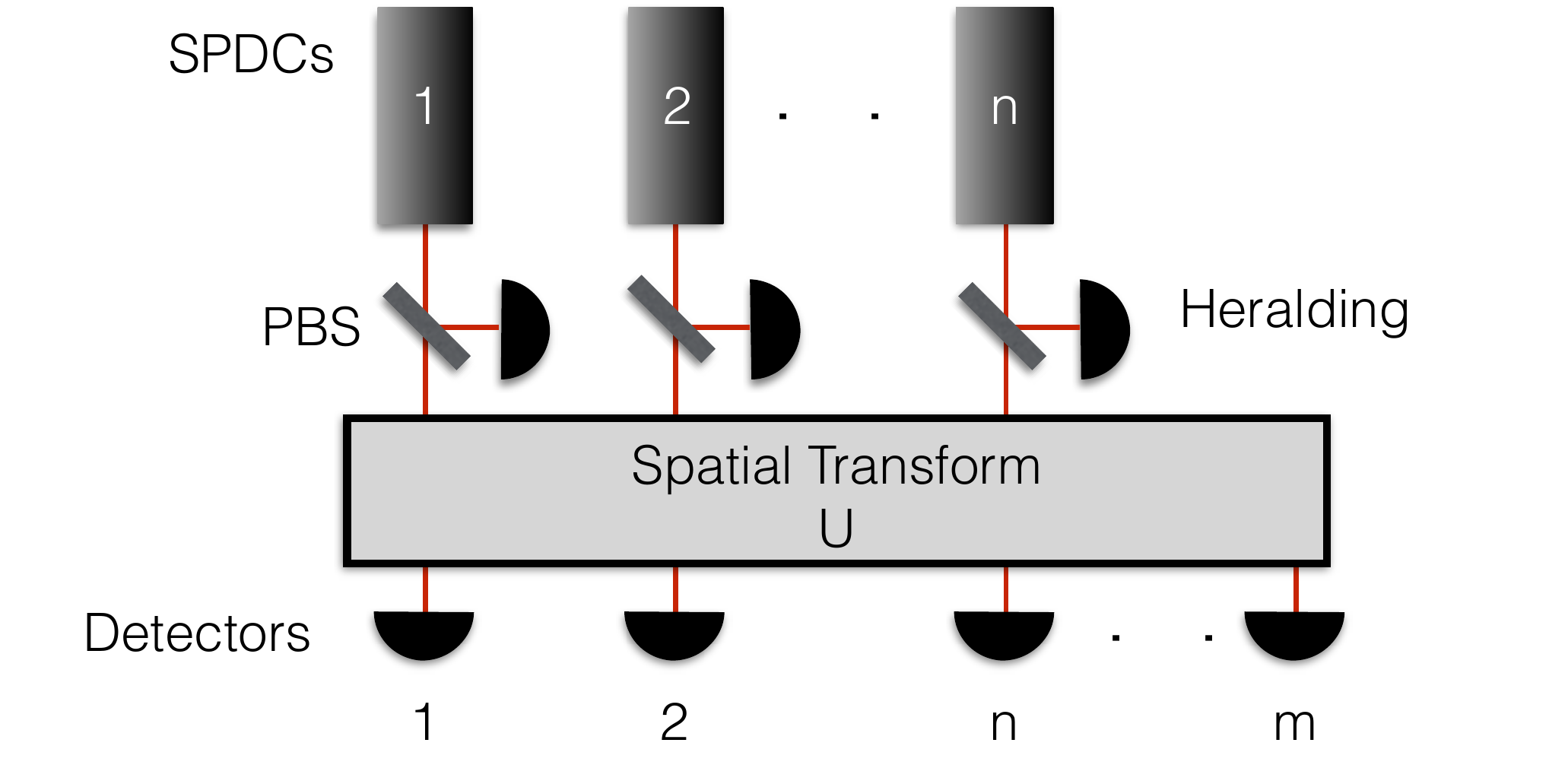}
    \includegraphics[width=0.5\textwidth, page=2]{Figs.pdf}
    \caption{Schematic of (a) spatial mode Boson Sampling (SMBS) (b) temporal mode Boson Sampling (TMBS)} 
    \label{scheme_conceptual}
\end{figure}

\begin{figure}[h]
    \includegraphics[width=0.5\textwidth, page=3]{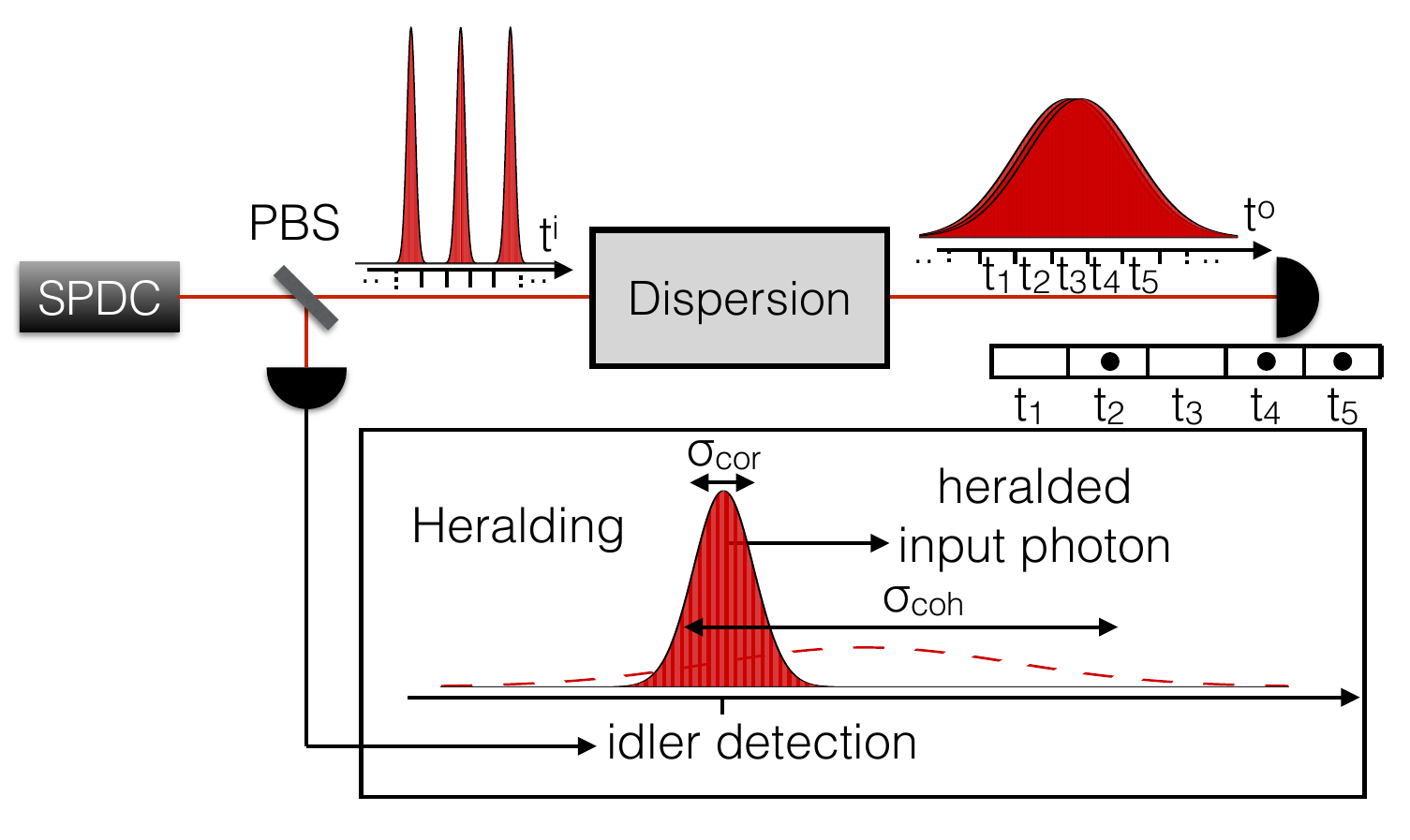}
    \caption{Schematic of TMBS with time-independent dispersion} 
    \label{schematic_TIdispersion}
\end{figure}

In conventional Boson Sampling schemes that use spatial modes (which we now refer to as `Spatial Mode Boson Sampling' or SMBS), multiple identical photons enter a high-dimensional transformation over spatial modes, such as a system of beamsplitters and phase shifters, while the output probability distribution is monitored with detectors at each of the output modes \cite{2013.Science.Spring-Walmsley.BS_photonic_chip, 2013.Science.Broome-White.BS_photonic_tunable, 2013.NPhoton.Tillmann-Walther.BS_silica_laser, 2013.NPhoton.Crespi-Sciarrino.BS_mmInterferometer, 2014.NPhoton.Carolan-Laing.BS_5photon, 2014.ArXix.Tillmann-Walther.BS_controllable_distinguishability, 2014.NPhoton.Spagnolo-Sciarrino.BS_Exp_val}, as shown in Fig.~\ref{scheme_conceptual} (a). Specifically, photons are injected into input modes $1....j...n$. The system then transforms the creation operator for input spatial mode $j$ as $\hat{a}_j^{\dagger} \rightarrow \sum_i U_{ij}\hat{a}^{\dagger}_i$, at which point photons in each mode are measured using single photon detectors \cite{2013.TheoryofComputing.Aaronson-Arkhipov.linear_optics_complexity, 2013.Science.Spring-Walmsley.BS_photonic_chip, 2013.Science.Broome-White.BS_photonic_tunable, 2013.NPhoton.Tillmann-Walther.BS_silica_laser, 2013.NPhoton.Crespi-Sciarrino.BS_mmInterferometer, 2014.NPhoton.Carolan-Laing.BS_5photon}. As we describe, Boson Sampling can analogously be performed in time by replacing \textit{spatial} mode $a_j$ with \textit{temporal} mode $a_{tj}$ (Fig.~\ref{scheme_conceptual} (b)) and by replacing the beamsplitter array with dispersion (Fig.~\ref{schematic_TIdispersion}).

SMBS entails several difficult challenges, that, as we show, favor temporal mode encoding. First, Boson Sampling requires an extremely large number of modes to be classically computationally difficult. Strictly speaking, the complexity argument for Boson Sampling assumes that, if $n$ is the number of photons in the system and $m$ is the number of modes, $m \geq \Omega(n^5\log^2n)$. Although Aaronson and Arkhipov have conjectured that the complexity arguments still hold when $m = O(n^2)$ \cite{2013.TheoryofComputing.Aaronson-Arkhipov.linear_optics_complexity}, the number of modes is still large: e.g., even with $m = n^2$ and $n = 30$, the interferometer would require 900 modes. Furthermore, the experiment would require 900 detectors and, if the photons came from heralded sources, 900 sources \cite{2014.PRL.Lund-Ralph.bs_gaussian}. To date, experimental demonstrations of SMBS have been limited to 5 photons in 21 modes \cite{2014.NPhoton.Carolan-Laing.BS_5photon}. The number of modes in temporal mode Boson Sampling (TMBS) can be increased simply by increasing the dispersion. Even with 10000 ps/nm dispersion, which can be achieved with off-the shelf components, and 100 ps detector jitter, which can be routinely achieved with silicon avalanche photodiodes or with superconducting nanowire single photon detectors \cite{2015.NatureComm.Najafi-Englund.OnChipSNSPD}, the number of modes in TMBS is orders of magnitude higher than SMBS. In principle, the experiment can be implemented in a single fiber with only a single photon source and only two detectors: one to herald input photons and one to detect the output state, regardless of the number of interfering photons in the system. Given the dead time $t_{dt}$ of single photon detectors, the output may have to be split between a larger number of detectors. However, in general, the number of detectors is smaller than required in SMBS \footnotemark[1]. 

Furthermore, uncertainty in the time when photons are injected into different modes leads to distinguishability and loss of boson interference.  This is a particular problem in SMBS with heralded sources based on spontaneous parametric down conversion (SPDC). Most SMBS experiments to date have relied on downconverted photons. Temporal or spectral filtering could improve the interference, but at an exponential loss in multi-photon throughput. In TMBS, the lack of control over the input time of our photons only corresponds to a lack of control over the choice of our input modes; however, as long as the input modes are known, this does not affect the ability to perform Boson Sampling \cite{2014.PRL.Lund-Ralph.bs_gaussian}. 

Previous proposals have considered temporal modes for Boson Sampling \cite{2014.ArXiv.Motes-Rhode.timeBS_loop, 2013.PRL.Humphreys-Walmsley.SingleSpatialModeLOQC} but they relied on temporarily converting temporal modes to spatial modes and then mixing the modes with beamsplitter operations. Hence, increasing the number of output modes ($m \gg n $) requires a large number of effective beamsplitter operations. They also require active elements that operate on a picosecond time scale. Furthermore, since they are based on the interference of narrow photon packets, they suffer from the same issues with temporal mismatch as SMBS. In TMBS, detector jitter can limit the accuracy with which the input mode can be heralded but this limitation can be overcome by using large dispersion. \newline

\noindent \textit{Time-independent dispersion - } We first consider the simplest case of Boson Sampling in time with identically shaped input photons and time-independent dispersion (Fig.~\ref{schematic_TIdispersion}). If we use an SPDC source with idler photons heralded at times $t_j$ and signal photons used as input photons, the input state is given by $\ket{\Psi_{in}} = \left(\prod_{j} \hat{a}^\dagger_{Aj}\right) \ket{0}$ where $\ket{0}$ is the multimode vacuum state and $\hat{a}^{\dagger}_{Aj} \equiv \int_{-\infty}^{\infty}\textrm{d}t\ \hat{a}^{\dagger}(t) A(t - t_j)$ represents the creation operator for the input state centered at $t_j$. $\hat{a}^{\dagger}(t)$ is the creation operator for time $t$ and $\omega_0$ is the central frequency of the input photons. We assume that the photon state after heralding of the idler is a pure state of the form $A(t-t_j)$. However, a realistic detector projects the signal photon into a mixed state with $t_j$ varying over the timescale of the detector jitter; the effect of this temporal mismatch can be made negligible with large dispersion \footnotemark[1]. 

$\hat{a}^\dagger_{Aj}$ can be expanded in the frequency domain as $\hat{a}^{\dagger}_{Aj} = \int_{-\infty}^{\infty}\textrm{d} \omega \hat{a}^{\dagger}(\omega) \mathcal{F} \{ A(t - t_j) \}$ where $\hat{a}^{\dagger}(\omega)$ is the creation operator for frequency $\omega$ and $\mathcal{F} \{ A(t - t_j) \}$ is the Fourier transform of $A(t - t_j)$. After passing through a dispersive element with dispersion relation $\beta(\omega)$ and length $L$, frequency components at $\omega$ are multiplied by a factor $\textrm{e}^{-i\phi(\omega)}$ where $\phi(\omega) \equiv \beta(\omega)L$. The wavefunction of the multi-photon system is then given by $\ket{\Psi_{out}} = \left(\prod_{j} \hat{b}^{\dagger}_{j}\right) \ket{0}$ where $\hat{b}^{\dagger}_j \equiv \int_{-\infty}^{\infty}\textrm{d} \omega \hat{a}^{\dagger}(\omega) \mathcal{F} \{ A(t - t_j) \}\textrm{e}^{-i\phi(\omega)}$. Going back to the time domain, $\hat{b}^{\dagger}_j = \int_{-\infty}^{\infty}\textrm{d} t \hat{a}^{\dagger}(t) U(t,t_j)$ with 

\begin{equation}
U(t,t_j) = A(t - t_j) * \mathcal{F}^{-1} \{ \textrm{e}^{-i\phi(\omega)} \}
\label{Uttj}
\end{equation}

\noindent where `$*$' is the convolution operator. 

If dispersion parameters are chosen such that $U(t,t_j)$ does not change appreciably when $t$ varies in a window of width $t_s$ \footnotemark[1], the modes can be discretized so that the transformation is well approximated by $\hat{a}^{\dagger}_{Aj} \rightarrow \sum_i U_{ij}\hat{a}^{\dagger}_{ti}$ where $\hat{a}^{\dagger}_{ti}$ represents the creation operator at the discretized time step near $t_i$, i.e. $\hat{a}^{\dagger}_{tj} = \int_{t_j}^{t_j+t_s} \textrm{d}t \hat{a}^{\dagger}(t)/\sqrt{t_s}$. We can then write the transformation as $\hat{a}^{\dagger}_{Aj} \rightarrow \sum_i U_{ij}\hat{a}^{\dagger}_{ti} = \int_{-\infty}^{\infty} \textrm{d}t U(t,t_j)\hat{a}^{\dagger}(t_i)$. If we assume that $U(t,t_j)$ is approximately constant for a small time step $t_s$, we can approximate $U_{ij}\hat{a}^{\dagger}_{ti} \approx U(t_i,t_j)\int_{t_j}^{t_j+t_s} \textrm{d}t \hat{a}^{\dagger}(t)$. Hence, we have $U_{ij} = \sqrt{t_s} U(t_i,t_j)$. $\hat{a}^{\dagger}_{Aj} \rightarrow \sum_i U_{ij}\hat{a}^{\dagger}_{ti}$  where

\begin{equation}
U_{ij}  =  \left[\sqrt{t_s} A(t - t_j) * \mathcal{F}^{-1} \{ \textrm{e}^{-i\phi(\omega)} \}\right]_{t = t_i}
\label{Uij_fixed}
\end{equation}

Eq.~\ref{Uij_fixed} shows the class of unitary transformations from which we can sample using time-independent dispersion. The unitary is band-diagonal because of the time-invariant nature of the system. Classical algorithms exist for the computation of the permanent of banded matrices with a banded inverse which is polynomial in the size of the matrix but exponential in the number of bands \cite{2012.ArXiv.Temme-Wocjan.PermBlockFact}. The inverse of a unitary banded matrix is banded. However, because the number of bands is extremely large and the number of bands/dispersion is increased with the number of photons in order to limit the effect of jitter \footnotemark[1], the problem is still expected to be computationally hard.

The shape of the input pulses $A(t)$ is incorporated into the unitaries in Eq.~\ref{Uij_fixed} because, unlike conventional unitary implementations, the input states and measurement have different bases; the input photons have shape $A(t)$ but the measurement is in the time basis (with eigenfunctions $\delta(t-t_j)$). The results of an experiment will be the same as a spatial unitary implementing Eq.~\ref{Uij_fixed}. Imagine a fictitious experiment where the input photons are $\delta(t-t_j)$. They then go through a unitary $U_1$ which puts them in a superposition of the form $A(t)$ (physically, this is the state of the photons going into the dispersion). The photons then go through another unitary $U_2$ which is the dispersion and the total unitary implemented by this system is $U_1U_2$. The output of this system will be the same as our scheme. Although a wavefunction $\delta(t-t_j)$ is unphysical, the detector sees the same output state as if the operator $U_1U_2$ was applied to photons of the form $\delta (t-t_j)$.

It is possible to sample from a larger class of unitaries by shaping the temporal form of the input photons. Methods for shaping single photons with arbitrary amplitude and phase in time have been proposed \cite{2010.PhysRevA.Kalachev.pulse_shaping} and an experimental demonstration of shaping the spatial waveform of single photons had been reported \cite{2011.OpticsLett.Koprulu-Kumar.pulseshape_spatial}. With pulse-shaping, the input waveform $A(t-t_j)$ is replaced by a more general set of functions $A_j(t)$ so that the accessible set of unitaries becomes

\begin{equation}
U_{ij}  =  \left[\sqrt{t_s} A_j(t) * \mathcal{F}^{-1} \{ \textrm{e}^{-i\phi(\omega)} \}\right]_{t = t_i}
\label{Uij_fixed_pulseshape}
\end{equation}

If it were possible to choose any set of functions $A_j(t)$, then the unitary could be chosen column by column using $A_j$ and simply detecting the photons without any dispersion would be equivalent to Boson Sampling. However, it is experimentally challenging to prepare multiple overlapping photons with a specific waveform. Hence, for realistic implementation, the photon wave-packets should be separated in time. This limits the possible unitaries represented by Eq.~\ref{Uij_fixed_pulseshape}. 

Although we have no proof of the hardness of sampling from such a unitary, there is to our knowledge no reported efficient classical algorithm for sampling from a general unitary of this form. Sending a train of photons through a time-independent dispersion could allow for a Boson Sampling experiment with more photons and more modes than can be currently achieved with SMBC.

An arbitrary functional form for the dispersion $\phi(\omega)$ can be obtained by using approaches used in optical functional design \cite{OFDAsignalprocapproach} and femtosecond pulse-shaping \cite{2000.RevSciInst.Weiner.femtopulseshape}. There are even commercial products for implementing arbitrary dispersion used for pulse-shaping in telecommunication \footnotemark[2].

A central feature in spatial boson collision experiments is the bunching of bosons in the output modes i.e., Hong-Ou-Mandel interference \cite{1987.PRL.Hong-Mandel.HOM}. An analogous feature appears in the temporal modes. Consider a heralded input state of two photons, $a^\dagger(t-t_1^i) a^\dagger(t-t_2^i)\ket{0}$. If this state passes through group velocity dispersion in a fiber of length $L$: $\beta(\omega) = \beta_0 + \beta_1(\omega-\omega_0) + 1/2\beta_2(\omega - \omega_0)^2$, $\phi_j = \beta_j L, j \in \{0,1,2\}$. For simplicity, we have assumed a Gaussian input shape of the form

\begin{equation}
A(t) = \left(\frac{1}{\sigma_{cor}\sqrt{2\pi}} \right)^{1/2} \exp{\left[-\frac{t^2}{4\sigma_{cor}^2}\right]} \exp[i\omega_0t],
\label{Aint_in_1}
\end{equation}

\noindent where we have assumed that the correlation time $\sigma_{cor}$ of the photons from the heralded photon source is much shorter than the biphoton coherence time. SPDC photons are often approximated as Gaussians in time. However, the two photon interference effects would be visible with any shape in general. After passing through this dispersive element, the probability of detecting the photons at times $t^o_1$ and $t^o_2$ corresponds to the magnitude squared of a $2 \times 2$ permanent and assuming $\phi_2 \gg \sigma_{cor}^2$, the probability goes to zero when \footnotemark[1]

\begin{equation}
\Delta t^i\Delta t^o = (2m+1)\pi \phi_2
\label{2photon_order2_zerocond}
\end{equation}

\begin{figure}[h]

    \includegraphics[width=0.5\textwidth, page=4]{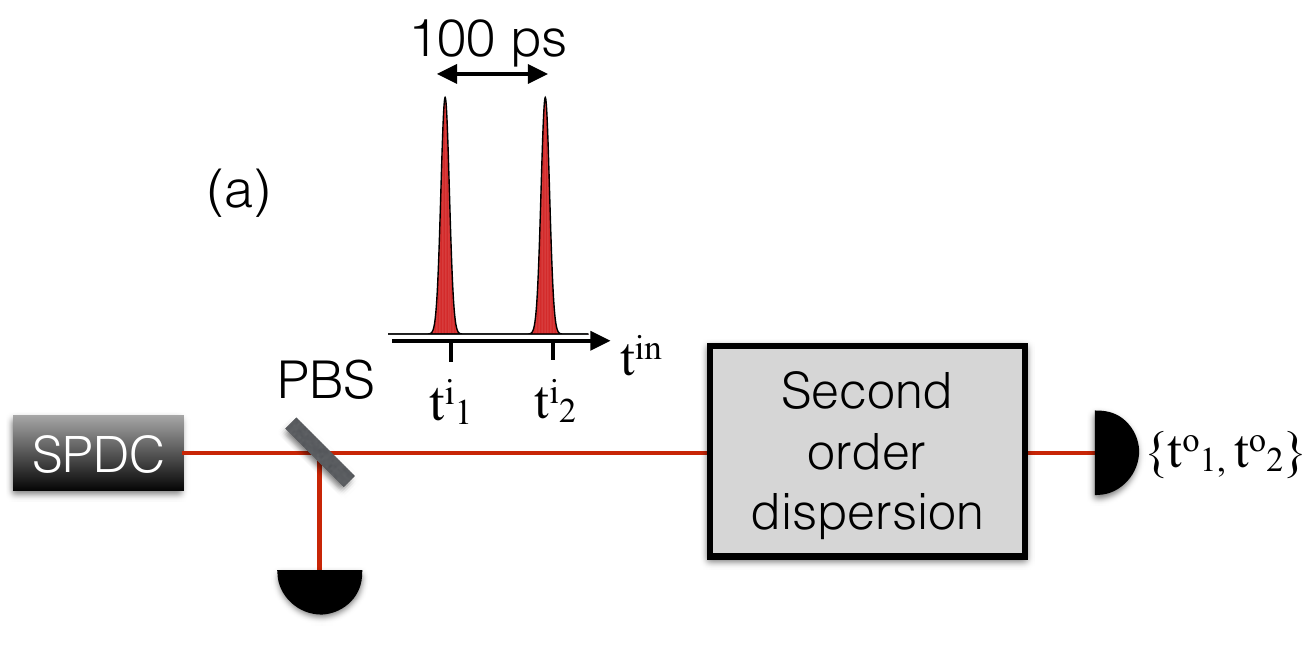}
    \includegraphics[width=0.4\textwidth, page=5]{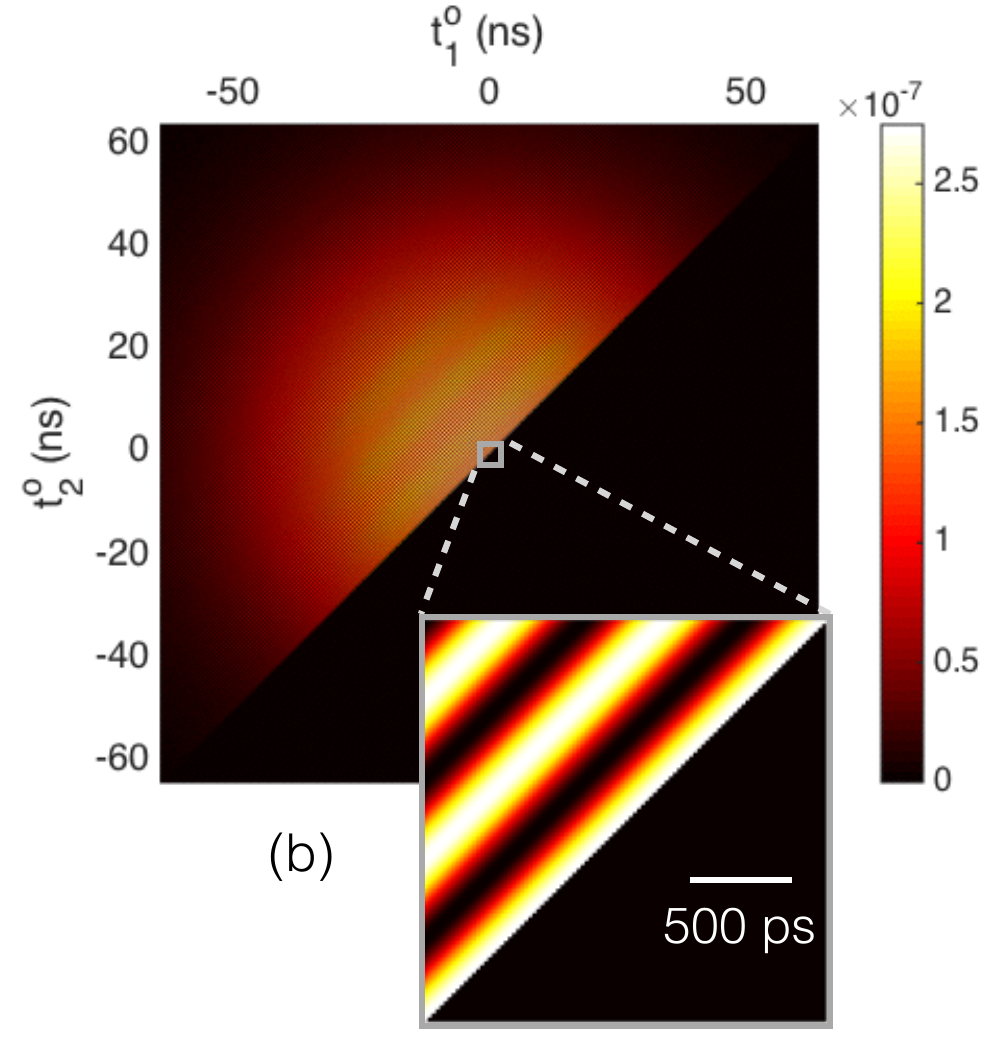}
    \caption{(a) Setup for seeing 'HOM-like' interference (b) The joint probability of detecting the first photon at $t^o_1$ and the second photon at $t^o_2$ when two input photons near t = 0 and separated by 100 ps are sent through a second order dispersive element. $\sigma_{cor}$ = 200 fs and the dispersive element has a GVD parameter of magnitude  $|D|$ = $2\pi c\phi_2/\lambda^2$ = 10000 ps/nm. $t_s = 10$ ps}
         \label{2photon_order2} 
\end{figure}

In Fig.~\ref{2photon_order2}, the joint probability of observing a photon at $t^o_1$ and $t^o_2$ is plotted when two photons with $\sigma_{cor}$ = 200 fs and a Gaussian temporal waveform centered at $t^i_1 = -\phi_1$ and $t^i_2 = -\phi_1 + 100$ ps are sent through a dispersive element with a GVD parameter of magnitude  $|D|$ = $2\pi c\phi_2/\lambda^2$ = 10000 ps/nm. The assumption $\phi_2 \gg \sigma_{cor}^2$ has not been used. At the time-scale of a few nanoseconds, the output resembles a Gaussian pulse centered at $t^o = 0$, as would be expected from single photon input at $t^i = -\phi_1$. However, at the time scale of hundreds of picoseconds, two-photon interference effects can be seen in clear dips in the two-photon output probability appear, as predicted by Eq.~\ref{2photon_order2_zerocond}. The plot has been discretized with $t_s = 10$ ps. We chose a low value of $t_s$ here to clearly show the shape of our interference pattern; such timing resolution is not necessarily required to resolve the two-photon interference pattern. Experimentally, these dips would be easily resolved using detectors with a jitter of 100 ps. The predicted correlation measurement is provided in Fig.~1 
of the supplemental material \footnotemark[1]. The increase in the size of each bin would also increase the probability of detecting photons in each time bin by a factor of 100. \newline

\begin{figure}[h]
    \includegraphics[width=0.5\textwidth, page=9]{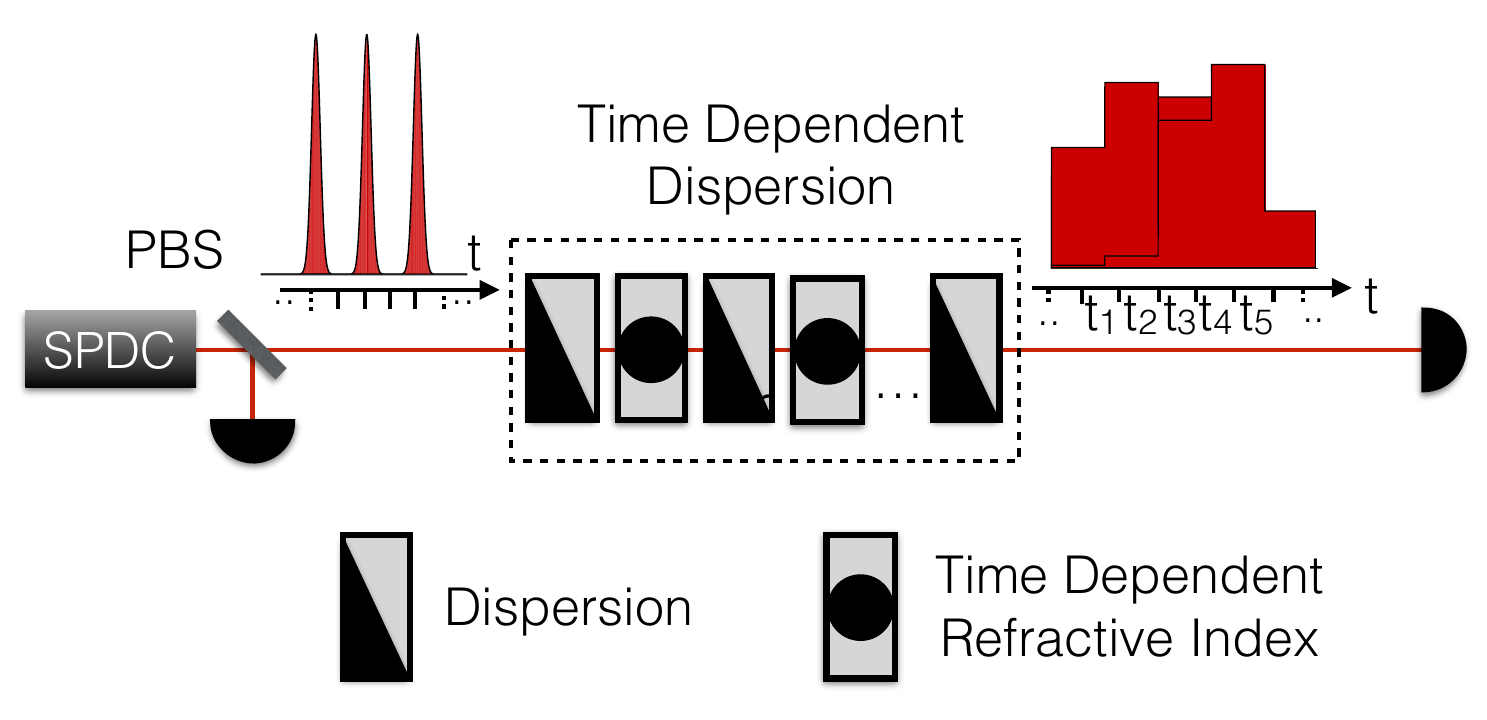}
    \caption{Schematic of TMBS with time-dependent dispersion} 
    \label{schematic_TDdispersion}
\end{figure}

\noindent \textit{Time-dependent dispersion - } With time-dependent dispersion, it is possible to implement any arbitrary unitary transformation on temporal modes.

Using the time projection operator $\hat{t}_p = \int \textrm{d}t' \ t \ket{t'} \bra{t'}$ and the frequency projection operator $\hat{\omega}_p = \int \textrm{d}\omega \ \omega \ket{\omega} \bra{\omega}$ which satisfy $[\hat{t}_p, \hat{\omega}_p] = i/2$ (similar to $\hat{x}$ and $\hat{p}$ in \cite{1999.PRL.Lloyd-Braunstein.CVQC, 2011.PRL.Sefi-Loock.DecomposeArbitraryCVUnitary}), an arbitrary continuous variable unitary on temporal modes can be written as $\textrm{e}^{-if(\hat{t}_p, \hat{\omega}_p)}$  where $f(\hat{t}_p, \hat{\omega}_p) = \int \textrm{d}t H(\hat{t}_p, \hat{\omega}_p, t)$. $H(\hat{t}_p, \hat{\omega}_p, t)$ is the Hamiltonian of the system. It should be noted that $\hat{t}_p$ is the time projection operator corresponding to the photon's temporal waveform and not the time over which the system evolves. Hence, the variable of integration $t$ above is distinct from the operator $\hat{t}_p$ .

Such a Hamiltonian can be physically realized by making the elements of the dispersion time-dependent. In the previous section we discussed methods for implementing arbitrary dispersion which implements a Hamiltonian of the form $\phi(\hat{\omega}_p)$. Since photons reaching the dispersive element at different times see a different $\phi(\hat{\omega}_p)$, the Hamiltonian can be written as $H(\hat{t}_p, \hat{\omega}_p, t)$ where $H$ can be any real function. Hence, we get the desired unitary $\textrm{e}^{-if(\hat{t}_p, \hat{\omega}_p)}$ with time-dependent elements that allow for changes in the frequency spectrum of the pulses which was not possible with time-independent dispersion. 

Another option for realizing such a unitary would be to cascade elements that implement time-independent dispersion and a time-dependent refractive index. The Hamiltonians corresponding to dispersion and time dependent refractive index are $\phi(\hat{\omega}_p)$ and $g(\hat{t}_p)$ respectively. 

Following previous work on realizing arbitrary Hamiltonians for continuous variable systems \cite{1999.PRL.Lloyd-Braunstein.CVQC, 2011.PRL.Sefi-Loock.DecomposeArbitraryCVUnitary}, one can construct Hamiltonians of the form $[g_1(\hat{t}_p), \phi_1(\hat{\omega}_p)] + [g_2(\hat{t}_p), [g_3(\hat{t}_p), \phi_3(\hat{\omega}_p)]] + .....$ using the properties 

\begin{eqnarray}
\textrm{e}^{-i\hat{A}\delta t}\textrm{e}^{-i\hat{B}\delta t}\textrm{e}^{i\hat{A}\delta t}\textrm{e}^{i\hat{B}\delta t} &=& \textrm{e}^{[\hat{A}, \hat{B}] \delta t^2} + O(\delta t^3) \label{commutehamiltonial} \\
\textrm{e}^{i\hat{A} \delta t/2}\textrm{e}^{i\hat{B} \delta t/2}\textrm{e}^{i\hat{B} \delta t/2}\textrm{e}^{i\hat{A} \delta t/2} &=& \textrm{e}^{i(\hat{A} + \hat{B}) \delta t} + O(\delta t^3) \label{sumhamiltonian}.
\end{eqnarray}


The following Hamiltonians can be cascaded to generate a Hamiltonian that can be any polynomial in $\hat{t}_p$ and $\hat{\omega}_p$ \cite{1999.PRL.Lloyd-Braunstein.CVQC}: $\hat{t}_p$, $\hat{\omega}_p$, $\hat{t}_p^2 + \hat{\omega}_p^2$ and a Hamiltonian of the form $\hat{\omega}_p^n$ where $n \geq 3$. The Hamiltonians $\hat{\omega}_p$ and $\hat{t}_p$ are first order dispersion (inverse group velocity) and a linear varying refractive index. A high order Hamiltonian $\hat{\omega}_p^n$ can be realized with higher order dispersion. $\hat{t}_p^2 + \hat{\omega}_p^2$ can be realized with a combination of second order dispersion and a quadratically varying refractive index using Eq.~\ref{sumhamiltonian}. Although such a decomposition allows us to build arbitrary Hamiltonians with a number of elements which increase as a small polynomial in the number of photons \cite{Braunstein2005}, more efficient decompositions are often possible with fewer elements \cite{2011.PRL.Sefi-Loock.DecomposeArbitraryCVUnitary}. 

Furthermore, as opposed to the conventional construction of $f(\hat{x}, \hat{p})$ \cite{1999.PRL.Lloyd-Braunstein.CVQC} which uses low-order polynomials in $\hat{x}$ and $\hat{p}$ (high order polynomials require a non-linear medium), a unitary of the form $g(\hat{t}_p)$ or $\phi(\hat{\omega}_p)$ can be of any arbitrary functional form without requiring an explicit Kerr-type nonlinearity.

Given the ability to realize arbitrary continuous variable single particle unitaries over $\hat{t}_p$, any discrete unitary transformation can be implemented by making the transformation constant over the output time bins \footnotemark[1]. Such an experiment with multiple photons would be equivalent to Boson Sampling and if the discrete unitary is chosen with Haar measure, the results are believed to be classically intractable \cite{2013.TheoryofComputing.Aaronson-Arkhipov.linear_optics_complexity}.

In conclusion, we have introduced new methods of implementing unitary transformations on temporal modes based on dispersion and pulse shaping that require a much smaller number of sources and detectors and do not require a large system of beamsplitters. In principle, using only fixed dispersion, a single heralded source and two detectors, one can observe multi-photon interference and perform a Boson Sampling experiment for which no efficient classical algorithm is known, to our knowledge. By using time-dependent dispersion, it is possible to sample from arbitrary unitaries.

We would like to thank Scott Aaronson, Alex Arkhipox, Gian Guerreschi and Ish Dhand for helpful discussions. This work was supported by the AFOSR MURI program under grant number (FA9550-14-1-0052).

We welcome any suggestions on this paper. You may contact us directly or submit comments at \url{http://goo.gl/forms/4YA8xRBj9s} (you may submit comments anonymously).

\footnotetext[1]{See Supplemental Material}
\footnotetext[2]{See \url{https://www.finisar.com/optical-instrumentation}}

\bibliography{Boson_Sampling}

\end{document}


\title{Supplemental Material: Scalable Boson Sampling Schemes by interference in the Time and Frequency Bases using dispersion and pulse shaping}

\begin{abstract}

\end{abstract}

\maketitle

\section{Number of detectors}

In this section we show that the number of detectors in TMBS can be much smaller than SMBS

In Fig.~
3 in the main paper, if photon detection is binned in steps of 100 ps, there are 1934 modes. The number of modes is defined as the number of time bins within which the absolute value of the dispersed wavefunction is greater than $90 \%$ of the peak value. 

We assume that our detectors have a dead time of 1 ns and look at the failure rate of our boson sampling scheme with 2000 input and output modes and 30 photons when each photon output is passively and equally split between 30 detectors. We assume that each time bin on the heralding as well on the output detector bank is equally likely to receive a photon. We post-select on the cases where there are a total of 30 photons incident on each detector bank (as in conventional scattershot boson sampling). 

The scheme is considered a failure if two photons are incident on any detector within the dead time. Based on the Monte-Carlo simulation of the system, we find that the probability of failure is less than $10 \%$

The assumption of photons being equally likely to arrive at any bin is accurate for the heralding detector bank. For the detector bank which detects photons after going through the unitary, the assumption may lead to an underestimated failure rate. However, we can postselect on the number of photons detected after the unitary being equal to the number of heralded photons and hence a higher failure rate is tolerable. An accurate simulation of the failure rate with a 30 photon, 2000 mode system is expected to be close to the limit of current computing capabilities.

Hence, TMBS can allow for a 30 photon 2000 mode experiment with 60 detectors, whereas an equivalent SMBS experiment would require 4000 detectors. It is interesting to observe that for the same number of photons and dead time bins, the number of detectors required for TMBS  goes down with an increase in the number of modes since there is a smaller chance of detecting multiple photons within a dead time reduces. In SMBS, the number of detectors is equal to twice the number of modes.

\section{Error bounds}

We find bounds on the error in the sampling distribution due to detector jitter and discretization. $U$ is the ideal unitary that we wish to implement, $\tilde{U}$ is the unitary with errors and $\mathcal{D}_{{U}}$ and $\mathcal{D}_{\tilde{U}}$ are the corresponding probability distributions over outcomes.

It has been shown in \cite{2014.ArXiv.Arkhipov.BSerror} that if there are $n$ photons in the system,

\begin{equation}
\|{ \mathcal{D}_{\tilde{U}} - \mathcal{D}_{{U}} }\| \leq n\|{\tilde{U}-U}\|_{op}
\end{equation}

If the relative error of each matrix element has a upper bound of $R$ i.e. $|\tilde{U}_{ij}-U_{ij}| \leq R|U_{ij}|$ for all $i$, $j$, 

\begin{eqnarray}
\|{ \mathcal{D}_{\tilde{U}} - \mathcal{D}_{{U}} }\| &\leq& n\|{RU}\|_{op} \nonumber \\
\|{ \mathcal{D}_{\tilde{U}} - \mathcal{D}_{{U}} }\| &\leq& nR
\end{eqnarray}

\noindent where we have used the fact that U is unitary.

Hence, in order to have $\|{ \mathcal{D}_{\tilde{U}} - \mathcal{D}_{{U}} }\|= o(1)$, $R = o\big(\frac{1}{n}\big)$.  

\subsection{Error due to detector jitter}
We have shown that the in the case of input photons with a fixed shape $A(t-t_j)$ sent through a dispersion $\beta(\omega)L$, the resulting unitary sampled from is 

\begin{eqnarray}
U_{ij} &=& \sqrt{t_s}U(t_i,t_j)   \\
&=& \sqrt{t_s}A(t_i-t_j) * B(t_i) \nonumber \\
 &=& \sqrt{t_s}A(t_i) * B(t_i - t_j) \nonumber
\end{eqnarray}

\noindent where $B(t) = \mathcal F^{-1}\{\textrm{e}^{-i\phi(\omega)}\}$.

Due to detector jitter, there is an uncertainty in $t_j$. If the maximum timing error due to detector jitter is $t_e$,

\begin{eqnarray}
R &=& \left|\frac{A(t_i) * B(t_i-t_j + t_e) - A(t_i) * B(t_i-t_j)}{A(t_i) * B(t_i-t_j)}\right| \nonumber \\
&\approx& \left|\frac{A(t_i) * \dot B(t_i-t_j)}{A(t_i) * B(t_i-t_j)}t_e\right|
\end{eqnarray}

If we write the dispersion in the form 

\begin{equation}
\phi(\omega) = \phi' (\phi_s \omega)
\label{dispasbetascale}
\end{equation}

\noindent where the $\phi_s$ is used to scale dispersion. It can be seen that $\dot B(t_i-t_j)/B(t_i-t_j) = o(1/\phi_s)$. Hence, if $A(t)$ is chosen independent of $n$, $R = o(t_e/\phi_s)$. Therefore, in order to limit the error due to detector jitter, $t_e/\phi_s = o(1/n)$.

\subsection{Error due to discretization}

In the case of time independent dispersion, the wavefunction generated after dispersion is treated discretely in order to draw a parallel with the original formalism for Boson Sampling \cite{2013.TheoryofComputing.Aaronson-Arkhipov.linear_optics_complexity}. 

$U_{ij} = \sqrt{t_s}U(t_i, t_j)$. The relative error can be written as

\begin{eqnarray}
R &<& \textrm{max}_t\left\{\left|\frac{U(t, t_j) - U(t_i,t_j)}{U(t_i,t_j)}\right|\right\} \nonumber \\
&\approx& \textrm{max}_t\left\{\left|\frac{\dot U(t,t_j)}{U(t_i,t_j)}\right|\right\} t_s
\end{eqnarray}

 \noindent where $t \in [t_i - t_s/2, t_i + t_s/2]$. If we define $\zeta = \textrm{max}_t\{\dot U(t,t_j)/U(t_i,t_j)\}$, we can see that $R = o(\zeta t_s)$. As discussed previously, we have $\zeta = o(1/\phi_s)$. Hence, in order to limit the error, $t_s/\phi_s = o(1/n)$.
 
\section{Temporal analog of the HOM dip}

In this section, we derive the two-photon wavefunction obtained on passing two photons with Gaussian envelopes through second order dispersion which is used to derive Eq.~5 
in the paper.

Using Eq.~1 of the main paper, a single photon with temporal waveform $A(t) = \left(\frac{1}{\sigma_{cor}\sqrt{2\pi}} \right)^{1/2} \exp{\left[-\frac{t^2}{4\sigma_{cor}^2}\right]} \exp[i\omega_0t]$ on passing through second order dispersion results in the temporal waveform

\begin{eqnarray}
\tilde{A}(t) &=& (2\pi)^{-1/4} \sqrt{\frac{\sigma_{cor}}{\sigma_{cor}^2+i\phi_2/2}} \textrm{e}^{-i\phi_0} \textrm{e}^{i\omega_0t} \\
& &\exp \left[-\frac{(t-\phi_1)^2}{4(\sigma_{cor}^4+\phi_2^2/4)}\left( \sigma_{cor}^2 - i\frac{\phi_2}{2}\right)\right] \nonumber
\end{eqnarray}

Hence, for two sets of entangled photons generated from a heralded source with idlers are detected at times  $t^i_1$ and $t^i_2$, after passing through the dispersive element, the wavefunction is given by

\begin{eqnarray}
\ket{\Psi} = \int_{-\infty}^{\infty}\textrm{d}t'\int_{-\infty}^{\infty} \textrm{d}t'' & & \hat{a}^{\dagger}(t') \hat{a}^{\dagger}(t'')  \nonumber \\
& & \tilde{A}(t' - t^i_1)\tilde{A}(t'' - t^i_2)
\label{2photon_HOM1_cont}
\end{eqnarray}

Every combination of $a^\dagger(t_1)a^\dagger(t_2)$ is repeated twice under the integrals. The repetition can be removed by rewriting the expression as

\begin{eqnarray}
& &\ket{\Psi} =  \int_{-\infty}^{\infty}\textrm{d}t' \int_{t'}^{\infty} \textrm{d}t'' \hat{a}^{\dagger}(t') \hat{a}^{\dagger}(t'')  \frac{\sigma \textrm{e}^{i\omega_0(t'+t'')-i2\phi_0}}{\sqrt{2\pi}(\sigma_{cor}^2 + i\phi_2/2)}\nonumber \\
& &  \Big\{ 1 + \exp{\left[\frac{-\sigma_{cor}^2+i\phi_2/2}{4\sigma_{cor}^4 + \phi_2^2}\{2(t^i_1-t^i_2)(t'-t'')\}\right]}\Big\} \nonumber \\
& & \exp\Big[\frac{-\sigma_{cor}^2+i\phi_2/2}{4\sigma_{cor}^4 + \phi_2^2}\{(t' - t^i_1-\phi_1)^2 \nonumber \\
& & + (t'' - t^i_2-\phi_1)^2\}\Big] \ket{0}
\label{2photon_HOM2_cont}
\end{eqnarray}

assuming that the pulse broadening due to the dispersion is much greater than the correlation time of the photons from the entangled source ($\phi_2 \gg \sigma_{cor}^2$), this is reduced to 

\begin{eqnarray}
& &\ket{\Psi} =  \int_{-\infty}^{\infty}\textrm{d}t' \int_{t'}^{\infty} \textrm{d}t'' \hat{a}^{\dagger}(t') \hat{a}^{\dagger}(t'')  \frac{\sigma \textrm{e}^{i\omega_0(t'+t'')-i2\phi_0}}{\sqrt{2\pi}(\sigma_{cor}^2 + i\phi_2/2)}\nonumber \\
& &  \Big\{ 1 + \exp{\left[i\frac{(t^i_1-t^i_2)(t'-t'')}{\phi_2}\right]}\Big\} \nonumber \\
& & \exp\Big[\frac{-\sigma_{cor}^2+i\phi_2/2}{4\sigma_{cor}^4 + \phi_2^2}\{(t' - t^i_1-\phi_1)^2 \nonumber \\
& & + (t'' - t^i_2-\phi_1)^2\}\Big] \ket{0}
\label{2photon_HOM3_cont}
\end{eqnarray}

From the equation above, Eq.~5
from the paper follows immediately.

Using $\hat{a}^{\dagger}_{t_j} = \int_{t_j}^{t_j+t_s} \textrm{d}t \hat{a}^{\dagger}(t)/\sqrt{t_s}$ to discretize Eq.~\ref{2photon_HOM2_cont}, we get

\begin{equation}
\ket{\Psi} =  \sum_{t_1} \sum_{t_2> = t_1}  \hat{a}^{\dagger}_{t_1} \hat{a}^{\dagger}_{t_2}  \textrm{Per}(\textbf{M})\ket{0}
\label{2photon_HOM_discrete}
\end{equation}

where 

\begin{eqnarray}
M_{jk} &=& \left(\frac{t_s\sigma}{\sqrt{2\pi}(\sigma_{cor}^2 + i\phi_2/2)}\right)^{1/2}\textrm{e}^{i\omega_0t_j - \phi_0} \\
& & \exp\left[\frac{-\sigma_{cor}^2+i\phi_2/2}{4\sigma_{cor}^4 + \phi_2^2}(t_j - t^i_k-\phi_1)^2\right]  \nonumber
\label{2photon_HOM_M}
\end{eqnarray}

\begin{figure}[h]
    \includegraphics[width=0.5\textwidth, page=8]{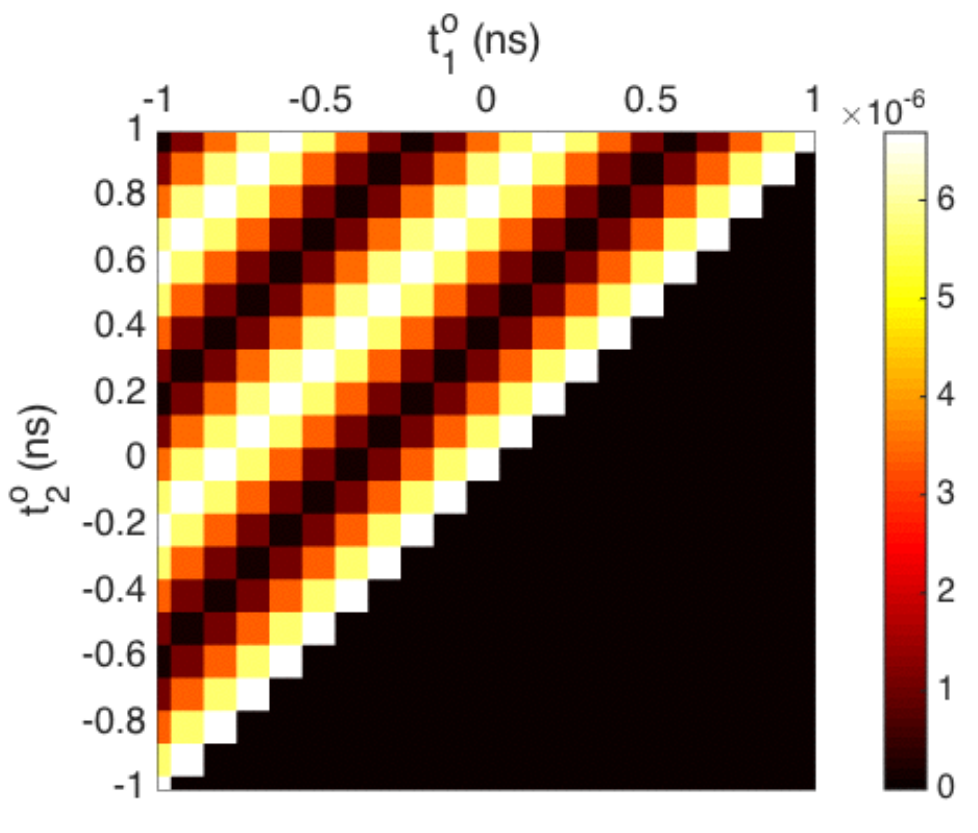}
    \caption{The joint probability of detecting the first photon at $t^o_1$ and the second photon at $t^o_2$ when two input photons near t = 0 and separated by 100 ps are sent through a second order dispersive element. $\sigma_{cor}$ = 200 fs and the dispersive element has a GVD parameter of magnitude  $|D|$ = $2\pi c\phi_2/\lambda^2$ = 10000 ps/nm. The probability has been binned into buckets of 100 ps which corresponds to a jitter achievable with currently available silicon and superconducting nanowire single photon detectors \cite{2015.NatureComm.Najafi-Englund.OnChipSNSPD}} 
    \label{2photon_order2_coarse}
\end{figure}

Hence, the probability of detecting photons at $t^o_1$ and $t^o_2$ is given by $[|\textrm{Per}(\textbf{M})|^2/(r_{in}!r_{out}!)]_{t_j = t^o_j}$ where $r_{in}(r_{out}) = 2$ if the  the input(output) photons are in the same mode and $1$ otherwise.

The joint probability of detecting two photons at $t^o_1$ and $t^o_2$ is plotted in Fig.~3 
in the paper with $t^i_1 = -\phi_1$, $t^i_2 = -\phi_1 + 100$ ps, $|D|$ = $2\pi c\phi_2/\lambda^2$ = 10000 ps/nm and $\sigma_{cor}$ = 200 fs. In the main paper the detection time has been binned in $10$ ps steps which is hard to achieve because of detector jitter. In Fig.~\ref{2photon_order2_coarse} here, the two photon interference pattern is visible even when the binning is increased to $100$ ps which is much easier to achieve experimentally.

\bibliography{Boson_Sampling}